\documentclass[prl,aps,twocolumn,amsmath,superscriptaddress,showpacs,floatfix]{revtex4-1}

\usepackage[euler]{textgreek}
\usepackage{graphicx}
\usepackage{SIunits}
\usepackage{color}
\usepackage{epstopdf}

\begin{document}

\title{Near-unity coupling efficiency of a quantum emitter to a photonic crystal waveguide}
\author{M.~Arcari}
\affiliation{Niels Bohr Institute, University of Copenhagen, Blegdamsvej 17, DK-2100 Copenhagen, Denmark}
\author{I.~S\"{o}llner}\email{sollner@nbi.ku.dk}
\affiliation{Niels Bohr Institute, University of Copenhagen, Blegdamsvej 17, DK-2100 Copenhagen, Denmark}
\author{A.~Javadi}
\affiliation{Niels Bohr Institute, University of Copenhagen, Blegdamsvej 17, DK-2100 Copenhagen, Denmark}
\author{S.~Lindskov~Hansen}
\affiliation{Niels Bohr Institute, University of Copenhagen, Blegdamsvej 17, DK-2100 Copenhagen, Denmark}
\author{S.~Mahmoodian}
\affiliation{Niels Bohr Institute, University of Copenhagen, Blegdamsvej 17, DK-2100 Copenhagen, Denmark}
\author{J.~Liu}
\affiliation{Niels Bohr Institute, University of Copenhagen, Blegdamsvej 17, DK-2100 Copenhagen, Denmark}
\author{H.~Thyrrestrup}
\affiliation{Niels Bohr Institute, University of Copenhagen, Blegdamsvej 17, DK-2100 Copenhagen, Denmark}
\author{E.~H.~Lee}
\affiliation{Center for Opto-Electronic Convergence Systems, Korea Institute of Science and Technology, Seoul, 136-791, Korea}
\author{J.~D.~Song}
\affiliation{Center for Opto-Electronic Convergence Systems, Korea Institute of Science and Technology, Seoul, 136-791, Korea}
\author{S.~Stobbe}
\affiliation{Niels Bohr Institute, University of Copenhagen, Blegdamsvej 17, DK-2100 Copenhagen, Denmark}
\author{P.~Lodahl}\email{lodahl@nbi.ku.dk} \homepage{http://quantum-photonics.nbi.ku.dk}
\affiliation{Niels Bohr Institute, University of Copenhagen, Blegdamsvej 17, DK-2100 Copenhagen, Denmark}

\date{\today}

\begin{abstract}
A quantum emitter efficiently coupled to a nanophotonic waveguide constitutes a promising system for the realization of single-photon transistors, quantum-logic gates based on giant single-photon nonlinearities, and high bit-rate deterministic single-photon sources.
The key figure of merit for such devices is the \textbeta-factor, which is the probability for an emitted single photon to be channeled into a desired waveguide mode. We report on the experimental achievement of $\beta = 98.43 \pm 0.04\%$ for a quantum dot coupled to a photonic-crystal waveguide, corresponding to a single-emitter cooperativity of $\eta = 62.7 \pm 1.5$. This constitutes a nearly ideal photon-matter interface where the quantum dot acts effectively as a 1D "artificial" atom, since it interacts almost exclusively with just a single propagating optical mode. The \textbeta-factor is found to be remarkably robust to variations in position and emission wavelength of the quantum dots. Our work demonstrates the extraordinary potential of photonic-crystal waveguides for highly efficient single-photon generation and on-chip photon-photon interaction.
\end{abstract}


\maketitle

The proposals of quantum communication \cite{bib:Bennett.1984} and linear-optics quantum computing \cite{bib:Knill.2001} have been major driving forces for the development of efficient single-photon (SP) sources {\cite{bib:Lee2011, bib:Claudon2010, bib:Gazzano.2012}}. Furthermore, the access to photon nonlinearities that are sensitive at the SP level {\cite{bib:Rice1988,bib:Chang2007}} would open for novel opportunities of constructing highly efficient deterministic quantum gates {\cite{bib:Chang2007, bib:Gao2008, bib:Shen2011, bib:Vuletic2013, bib:Hwang2009, bib:Turchette1995, bib:Peyronel2012}}. A single quantum emitter that is efficiently coupled to a photonic waveguide \cite{bib:Shen2005} would facilitate such a SP nonlinearity, enabling the realization of single-photon switches and diodes \cite{bib:Chang2007, bib:Gao2008, bib:Shen2011}, as well as serve as a highly efficient single-photon source. Waveguide-based schemes offer highly efficient and broadband channeling of SPs into a directly usable propagating mode where even the photon detection can be integrated on-chip \cite{bib:Reithmaier.2013}. The associated SP nonlinearity constitutes a very promising and robust alternative to the technologically demanding schemes based on the anharmonicity of the strongly-coupled emitter-cavity system \cite{bib:Englund.2012, bib:Volz.2012, bib:Kim.2013, bib:Birnbaum.2005}.

In the present work we consider a single quantum dot (QD) embedded in a photonic-crystal waveguide (PCW). The important figure of merit is the \textbeta-factor:
\begin{equation}
\beta = \frac{\Gamma_{\text{wg}}}{\Gamma_{\text{wg}}+\Gamma_{\text{rad}}+\Gamma_{\text{nr}}} = \frac{\Gamma_{\text{c}}-\Gamma_{\text{uc}}}{\Gamma_{\text{c}}},
\end{equation}
which gives the probability for a single exciton in the QD to recombine by emitting a single photon into the waveguide mode. $\Gamma_{\text{wg}}$ and $\Gamma_{\text{rad}}$ are the rate of decay {of the QD into either the guided mode or non-guided radiation modes,} whereas $\Gamma_{\text{nr}}$ denotes the intrinsic nonradiative decay rate of the QD. The \textbeta-factor is related to the single-emitter cooperativity  $\eta = \beta/(1-\beta)$.  \cite{bib:Tiecke.2014}
Experimentally, the \textbeta-factor can be obtained by recording the decay rate of a QD that is coupled to the waveguide $\Gamma_{\text{c}} = \Gamma_{\text{wg}} + \Gamma_{\text{rad}} + \Gamma_{\text{nr}}$ and the rate of an uncoupled QD $\Gamma_{\text{uc}} = \Gamma_{\text{rad}}+\Gamma_{\text{nr}}$ in the case where the difference between the total loss rates $(\Gamma_{\text{rad}}+\Gamma_{\text{nr}})$ of the two QDs is negligible.

Recent proposals have indicated that the \textbeta-factor in PCWs may approach unity \cite{bib:MangaRao.2007, bib:Lecamp.2007}. However, measuring a near-unity \textbeta-factor is experimentally challenging, because the reliable extraction of $\Gamma_{\text{uc}}$ is not straightforward. A proper measurement of the \textbeta-factor requires the precise determination of $\Gamma_{\text{uc}}$ for a QD that is coupled to the waveguide. In previous work \cite{ bib:Hansen.2008, bib:Dewhurst2010, bib:Thyrrestrup2010, bib:Hoang2012, bib:Laucht2012} $\Gamma_{\text{uc}}$ was estimated either from QDs spectrally tuned to the band gap of the photonic crystal, or from QDs positioned outside the waveguide region (for a more detailed discussion of the previous experimental methods, see \cite{bib:Supp}). The drawback of both approaches is that the modifications in the coupling to the non-guided modes from the presence of the waveguide are not accounted for, which may lead to largely incorrect estimates of the \textbeta-factor, as revealed by numerical simulations.

In the present work, the \textbeta-factor is experimentally determined by comparing the decay rate of a QD coupled to the waveguide mode $\Gamma_{\text{c}}$ to that of a very weakly coupled QD, which constitutes an upper bound of $\Gamma_{\text{uc}}$.
One of the key differences to previous work is that both $\Gamma_{\text{c}}$ and $\Gamma_{\text{uc}}$ are obtained by directly detecting the propagating waveguide mode, hence all measured QDs are spatially positioned in the waveguide. This guarantees that the spatial and spectral dependence of the coupling to radiation modes due to the presence of the PCW are correctly taken into account, which is essential for the analysis. The validity of our experimental method is confirmed by numerical simulations of the position and frequency dependence of $\Gamma_{\text{rad}}$ in the PCW structure. We experimentally demonstrate a SP channelling efficiency of $\beta = 98.43 \pm 0.04\%$ for a QD in a PCW, which significantly surpasses previously reported results exploiting atoms \cite{bib:Vetsch.2010, bib:Tiecke.2014}, nitrogen vacancy centres \cite{bib:Babinec.2010}, single molecules \cite{bib:Lee2011}, or quantum dots \cite{bib:Claudon2010, bib:Hansen.2008, bib:Akimov.2007} as the photon sources in photonic-waveguide structures. This corresponds to a single-emitter cooperativity $\eta = 62.7 \pm 1.5$, which surpasses by almost one order of magnitude previously reported values both with QDs \cite{bib:Hansen.2008} and atoms \cite{bib:Tiecke.2014}. Such a high coupling efficiency matches the level achievable with superconducting microwave circuits, widely considered one of the most mature platforms for scalable quantum-information processing available today, and will lead to novel opportunities for photonic quantum-information processing \cite{bib:Lodahl.2013}.

\begin{figure}[t!]
\includegraphics[width=\columnwidth]{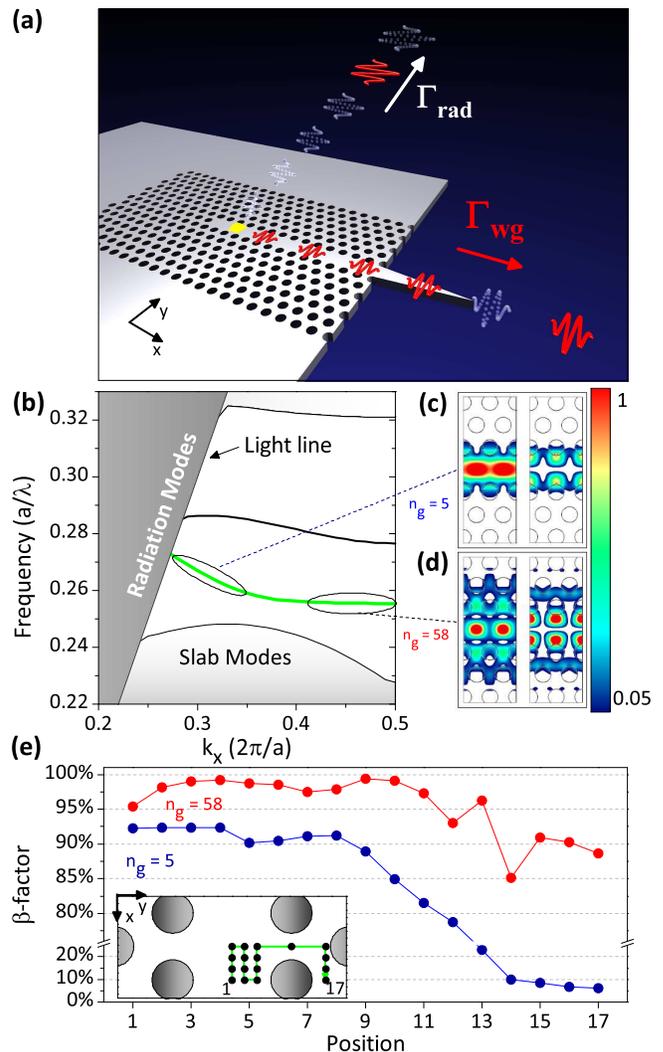}
\caption{(Color online).(a) Illustration of the device. A train of single-photon pulses (red pulses) are emitted from a triggered QD (yellow trapezoid). The photons are channeled with near-unity probability into the waveguide mode with a rate $\Gamma_{\text{wg}}$ while the weak rate $\Gamma_{\text{rad}}$ of coupling to radiation modes implies that only very few photons are lost. The guided photons can be efficiently extracted from the waveguide through a tapered mode adapter. (b) Projected TE band structure of the PCW studied in the experiments displaying even (green) and odd (black) waveguide modes in the band gap. In the experiment only the even mode is studied. The shaded area of the dispersion diagram corresponds to the continuum of radiation modes. (c)-(d) Electric field intensity spatial profiles $|E_x|^2$ (right) and $|E_y|^2$ (left) for two different spectral regions of the waveguide mode corresponding to two different group indices $n_{\text{g}}$. (e) Maximum \textbeta-factor of the two orthogonal dipoles at $n_{\text{g}} = 58$ (red) and $n_{\text{g}} = 5$ (blue) calculated at the positions indicated in the inset. The maximum \textbeta-factor is the quantity measured in decay-dynamics experiments.\label{fig:Figure1}}
\end{figure}

A near-unity \textbeta-factor PCW SP source is illustrated in Figure \ref{fig:Figure1}(a): a deterministic train of SPs in the waveguide can be obtained since the excited QD will emit a photon into the waveguide with probability $\beta$, while out-of-plane photon loss is strongly suppressed. High \textbeta-factors are achievable due to the combination of two effects: a broadband Purcell enhancement of the rate $\Gamma_{\text{wg}}$ of coupling into the waveguide and the strong suppression of the loss rate $\Gamma_{\text{rad}}$ due to the photonic-crystal membrane structure. Different physical systems have been proposed for obtaining a large \textbeta-factor: plasmonic nanowires rely on the Purcell enhancement thereby increasing $\Gamma_{\text{wg}}$  \cite{bib:Akimov.2007}, while dielectric nanowires \cite{bib:Claudon2010, bib:Babinec.2010} mainly suppress the coupling to radiation modes, i.e., decrease $\Gamma_{\text{rad}}$. In PCWs the beneficial combination of the Purcell enhancement of the PCW mode and the pronounced reduction of radiation modes enables a near-unity \textbeta-factor.

In PCWs the Purcell enhancement is proportional to the group index or slow-down factor $n_{\text{g}} = c/v_{\text{g}}$, where $c$ is the speed of light in vacuum. The group velocity of light, $v_{\text{g}}$, is the slope of the waveguide band, see Figure \ref{fig:Figure1}(b), {which decreases at the waveguide band edge}. {We have measured $n_{\text{g}}>$50} close to the band edge, leading to expected Purcell factors close to 10 \cite{bib:MangaRao.2007}. For the method used to extract the $n_{\text{g}}$, see \cite{bib:Supp}. Furthermore, the photonic-crystal band gap strongly inhibits the in-plane radiative loss rate of the dipole emitter, while total internal reflection limits the decay by out-of-plane radiation.

The calculated position dependent \textbeta-factor for a dipole emitter in proximity of the band edge ($n_{\text{g}}$ = 58) and close to the light line ($n_{\text{g}}$ = 5) is shown in Figure \ref{fig:Figure1}(e). The fraction of the emission coupled to the waveguide and to the radiation modes can be determined numerically and $\Gamma_{\text{wg}}$ and $\Gamma_{\text{rad}}$ are obtained by multiplying by the measured average radiative decay rate of QDs in a homogeneous medium $\Gamma_{\text{hom}} =$ 0.91$\pm$0.08 ns$^{-1}$. The measured average nonradiative decay rate is $\Gamma_{\text{nr}} =$ 0.030$\pm$0.018 ns$^{-1}$.
In the slow-light regime $(n_{\text{g}} = 58)$, \textbeta-factors exceeding $95 \%$ are predicted for dipole positions close to the field maxima of the waveguide mode (Figure \ref{fig:Figure1}(d)). In agreement with previous results \cite{bib:MangaRao.2007, bib:Lecamp.2007}, even outside the slow-light regime ($n_{\text{g}}$ = 5), \textbeta-factors exceeding 90\% are predicted for many positions in the waveguide, illustrating the extremely broadband coupling. The position dependence of the \textbeta-factor, that is displayed in Figure \ref{fig:Figure1}(e), is determined by the spatial mode profiles of the guided modes {shown} in Figure \ref{fig:Figure1}(c)-\ref{fig:Figure1}(d).

\begin{figure}[t!]
\begin{center}
\includegraphics[width=\columnwidth]{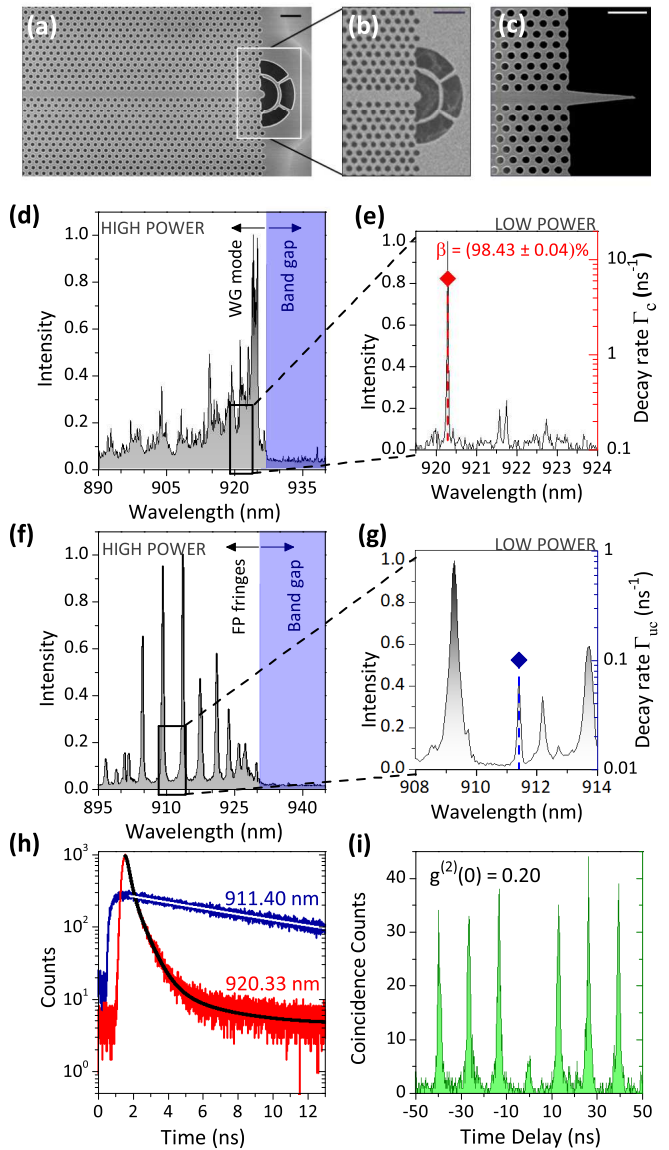}
\end{center}
\caption{(Color online). (a) Scanning-electron microscope image of a typical device. QDs are excited in an area centered on the PCW, and the resulting emission is collected either from (b) the grating or (c) the tapered mode adapter. The scale bars corresponds to 1 \textmu m. (d) High-power emission spectrum displaying the mode of the high-$n_{\text{g}}$ waveguide section for a grating structure with weakly reflecting ends. (e) Low-power spectrum with the corresponding decay rate $\Gamma_{\text{c}}$ and \textbeta-factor for a QD that is efficiently coupled to the waveguide. (f) High-power emission spectrum for a low-$n_{\text{g}}$ waveguide section with a highly reflective grating mode adapter. (g) Low power spectrum and decay rate $\Gamma_{\text{uc}}$ of a QD spectrally in-between two FP resonances. (h) Decay curves of QD-lines shown in (e) (red curve) and in (g) (blue curve). (i) Measured autocorrelation function of the QD in (e).}\label{fig:Figure2}
\end{figure}

We investigate light emission from a single layer of self-assembled InAs QDs embedded in a GaAs PCW (see \cite{bib:Supp} for sample description).
In order to efficiently collect the photons from the propagating waveguide mode, either second-order Bragg gratings \cite{bib:Faraon.2008} or inverse tapered mode adapters \cite{bib:Tran.2009} are used. A numerical study of the coupling efficiency of the two outcoupling methods is presented in \cite{bib:Supp}. Scanning-electron microscope images of typical devices are shown in Figure \ref{fig:Figure2}(a)-(c). Since the mode adapters are designed to work in the regime of low $n_{\text{g}}$, a transition region is introduced in the photonic crystal in order to couple from the high-$n_{\text{g}}$ waveguide mode into the low-$n_{\text{g}}$ mode \cite{bib:Hugonin.2007}. The length of the high-$n_{\text{g}}$ region varies between 5.1 and 8.3 \textmu m for different samples; a short sample length is chosen to eliminate the formation of Anderson-localized modes \cite{bib:Sapienza.2010}, which are detrimental for obtaining a high waveguide transmission. The averaged extinction length for light propagation in similar waveguides was measured to be $l\simeq$ 30 \textmu m \cite{bib:Smolka.2011}.

Figure \ref{fig:Figure2}(d) shows a high-power photoluminescence spectrum of the waveguide mode collected from the grating under non-resonant excitation, which is used to characterize the waveguide samples. The applied power  is approximately two orders of magnitude higher than the saturation power of single excitons, i.e., single QD lines can not be distinguished in this case. The spectrum displays a cut-off at 925 nm due to the waveguide band edge and a transmission bandwidth of 35 nm. Similar spectra were obtained when collecting the emission from the inverse tapers.
We also investigated waveguide structures where a change in parameters of the gratings led to high reflectivity and the formation of sharp Fabry-P\'{e}rot (FP) resonances within the waveguide bandwidth, as shown in Figure \ref{fig:Figure2}(f). In these structures the coupling to the waveguide mode for QDs spectrally positioned in-between two resonances is very weak, implying that {the measured $\Gamma_{\text{uc}}$ is close to the lower limit of $\Gamma_{\text{rad}}+\Gamma_{\text{nr}}$}.

Time-resolved photoluminescence spectroscopy is employed to characterize the dynamics of QDs in a 20 nm range, blue-detuned from the band edge cut-off. For each QD line, decay curves are measured at an excitation power level well below saturation, see Figure \ref{fig:Figure2}(h). For details on how the decay curves are modeled, see \cite{bib:Supp}.

\begin{figure}[t!]
\begin{center}
\includegraphics[width=\columnwidth]{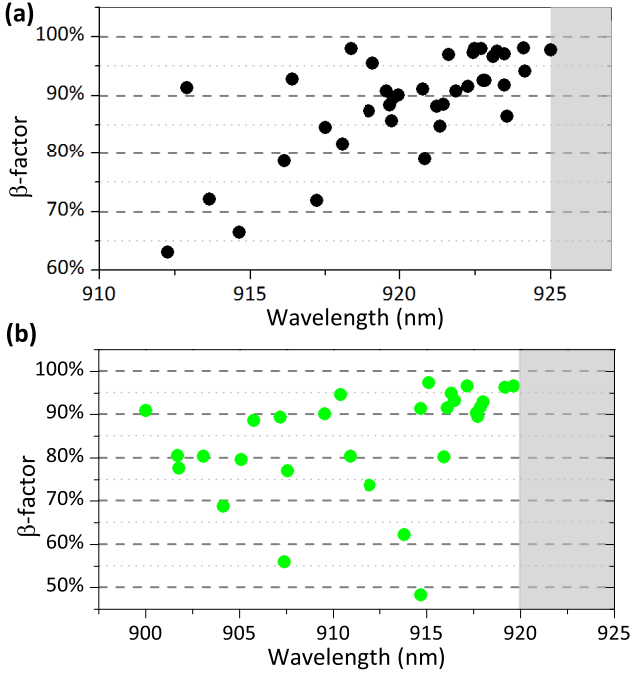}
\end{center}
\caption{(Color online). (a) \textbeta-factors measured on the grating samples and (b) on the samples with tapered waveguides. In both figures, the band gap region above the waveguide cutoff is indicated in grey.}\label{fig:Figure4}
\end{figure}

We extract Purcell-enhanced decay rates of up to $\Gamma_{\text{c}} = 6.28 \pm 0.15$ ns$^{-1}$ in the high-$n_{\text{g}}$ waveguide sections (Figure \ref{fig:Figure2}(e)) and inhibited decay rates down to $\Gamma_{\text{uc}} = 0.098 \pm 0.001$ ns$^{-1}$ between two FP resonances in the low-$n_{\text{g}}$ waveguide section (Figure \ref{fig:Figure2}(g)). This corresponds to $\beta = 98.43 \pm 0.04 \%$.
We have measured the \textbeta-factor for a total of 71 different QDs within a 20 nm range in both the samples with grating and taper mode adapters, and the results are shown in Figure \ref{fig:Figure4}. In both samples, the \textbeta-factors are above 90\% for most of the QDs in a 5 nm range close to edge of the band-gap region, and we measure \textbeta-factors above 90\% for QDs up to 20 nm spectrally detuned from the band edge, thus highlighting the robustness of these devices. These results agree very well with the theoretical predictions of Figure \ref{fig:Figure1}(e), which shows that QDs with \textbeta-factor above 90\% can be expected even at $n_\text{g} =5$ for specific positions in the waveguide. In the following we will focus on the highest \textbeta-factors found for QDs in the proximity of the band edge.
The SP nature of the emission lines is confirmed by recording the normally ordered second order intensity correlation function $g^{(2)}(\tau) = \langle :\hat{I}(t) \hat{I}(t+\tau): \rangle / \langle \hat{I}(t)\rangle^2$ under pulsed excitation. An example of a measurement for the highest $\beta$-factor QD of Figure \ref{fig:Figure2}(e) is shown in Figure \ref{fig:Figure2}(i), where $g^{(2)}(0) = 0.20$ is found at 0.63 of saturation power. For further details about the analysis of $g^{(2)}(\tau)$, see \cite{bib:Supp}. Even stronger antibunching has been observed for QDs in spectrally very clean regions of the waveguide mode reaching $g^{(2)}(0) < 0.05$ at excitation powers below the saturation level.

\begin{figure}[t!]
\begin{center}
\includegraphics[width=\columnwidth]{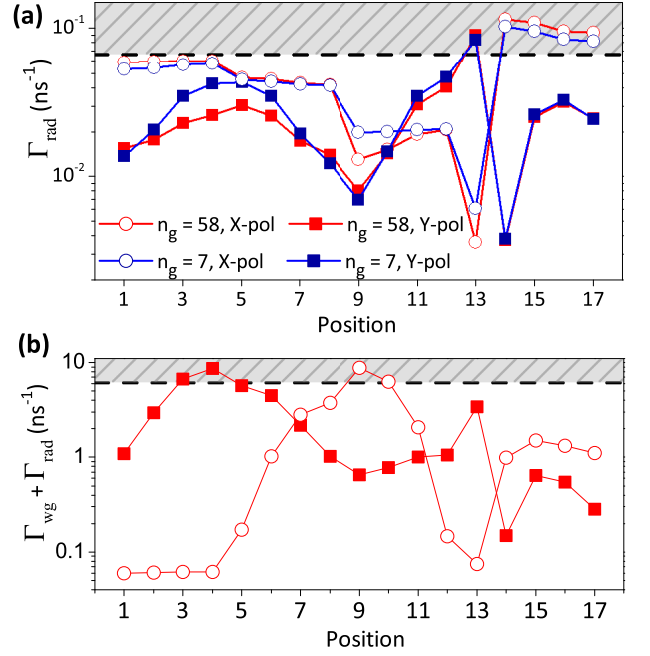}
\end{center}
\caption{(Color online). (a) Position-dependent radiative loss rate for a y-polarized (squares) or x-polarized (circles) dipole emitting at either $n_{\text{g}} = 58$ in the high-$n_{\text{g}}$ waveguide section (red data) or $n_{\text{g}} = 7$ in the low-$n_{\text{g}}$ waveguide section (blue data). The shaded region indicates where the predicted $\Gamma_{\text{rad}}$ is above the experimentally extracted value of $0.063 \pm 0.008$ ns$^{-1}$.
(b) Calculated radiative rate as a function of position for a dipole aligned along the y-direction (squares) or the x-direction (circles) and emitting at $n_{\text{g}} = 58$. The corresponding emitter positions are indicated in the inset of Figure \ref{fig:Figure1}(e). The shaded region indicates where the predicted decay rate is above the experimentally extracted value of $6.28 \pm 0.15$ ns$^{-1}$ for the high-\textbeta-factor QD.}\label{fig:Figure3}
\end{figure}

The applied method for extracting the \textbeta-factor by comparing $\Gamma_{\text{c}}$ and $\Gamma_{\text{uc}}$ of two different QDs is valid since the variation of the total loss rate $\Gamma_{\text{rad}} + \Gamma_{\text{nr}}$ between different QDs is small.
Indeed, the variations in $\Gamma_{\text{nr}}$ over the wavelength range of relevance can be neglected \cite{bib:Johansen.2010} while the spatial variation of $\Gamma_{\text{rad}}$ has been calculated as  shown in Figure \ref{fig:Figure3}(b) for an efficiently ($n_{\text{g}} =58$) and weakly coupled ($n_{\text{g}} =7$) QD. The chosen values of $n_{\text{g}}$ correspond to the cases of the QDs shown in Figure \ref{fig:Figure2}(e),\ref{fig:Figure2}(g). For most positions, $\Gamma_{\text{rad}}$ is found to be lower than the experimental estimate based on $\Gamma_{\text{rad}}=\Gamma_{\text{uc}}-\Gamma_{\text{nr}}$ (indicated by the dashed line in Figure \ref{fig:Figure3}(a)), which is expected since residual coupling to the waveguide will increase the rate.
At a few spatial positions in Figure \ref{fig:Figure3}(a), the predicted $\Gamma_{\text{rad}}$ is found to be higher than the experimental rate. However, as displayed in Figure \ref{fig:Figure3}(b) these are not the positions where the large Purcell-enhanced decay rates observed in the experiment appear (experimental values indicated by the dashed line in Figure \ref{fig:Figure3}(b)), and we can thus exclude that the QDs with highest \textbeta-factor are positioned here, implying that the record-high \textbeta-factors constitute conservative estimates.

In conclusion, we have for the first time unambiguously measured a near-unity \textbeta-factor in a PCW, and have shown the robustness of $\beta$ with respect to the wavelength and the spatial position of the emitter in the PCW. This result proves the high potential of PCWs to achieve strong on-chip light-matter coupling between a QD and a propagating mode, which opens the way for the realization of transistors and gates for deterministic quantum information processing. To this end highly coherent photon emission is required, and decoherence processes have been found to be slow for quasi-resonant excitation of QDs in photonic crystals \cite{bib:Laurent.2005, bib:Madsen.2014} and could be further extended by applying true resonant excitation \cite{bib:He2013, bib:Mathiasen.2012,bib:Nguyen.2011}. Furthermore, a highly-efficient SP source can be built by optimizing the outcoupling mode adapters of our structures, making the source immediately applicable for linear-optics quantum-computing experiments.

We acknowledge fruitful discussion with Anders S. S\o rensen. We thank Sara Ek for her input to the design of the tapered structures. We gratefully acknowledge financial support from the Danish Council for Independent Research (Natural Sciences and Technology and Production Sciences) and the European Research Council (ERC consolidator grant ALLQUANTUM).

I. S. and M. A. contributed equally to this work.

\clearpage
\newpage
\onecolumngrid
\renewcommand{\figurename}{FIG. S}
\section*{Supplementary material}
\subsection*{A. Sample design and extraction of the group index}

The samples consist of 160 nm thin GaAs (refractive index $n=3.45$) photonic-crystal membranes with lattice constants of either $a = 238$ nm or $a = 235$ nm and respective hole radii of $r = 0.29 a$ or $r = 0.31 a$. A waveguide is created by leaving out a single row of holes. A single layer of InAs self-assembled QDs, with a density of $\sim${100} {\textmu m}$^{-2}$ and an inhomogeneously broadened spectrum spanning the range of 920$\pm$25 nm, was grown by molecular beam epitaxy in the center of the membrane. The photonic crystal structures were defined on a layer of resist (ZEP 520A) by electron-beam lithography and then transferred into the GaAs layer by using a chlorine based coupled plasma reactive-ion etching. Suspended membrane structures were formed by removing a 1-\textmu m-thick AlGaAs sacrificial layer underneath the GaAs layer with hydrofluoric acid vapour.

\begin{figure}
\begin{center}
\includegraphics[scale = 1]{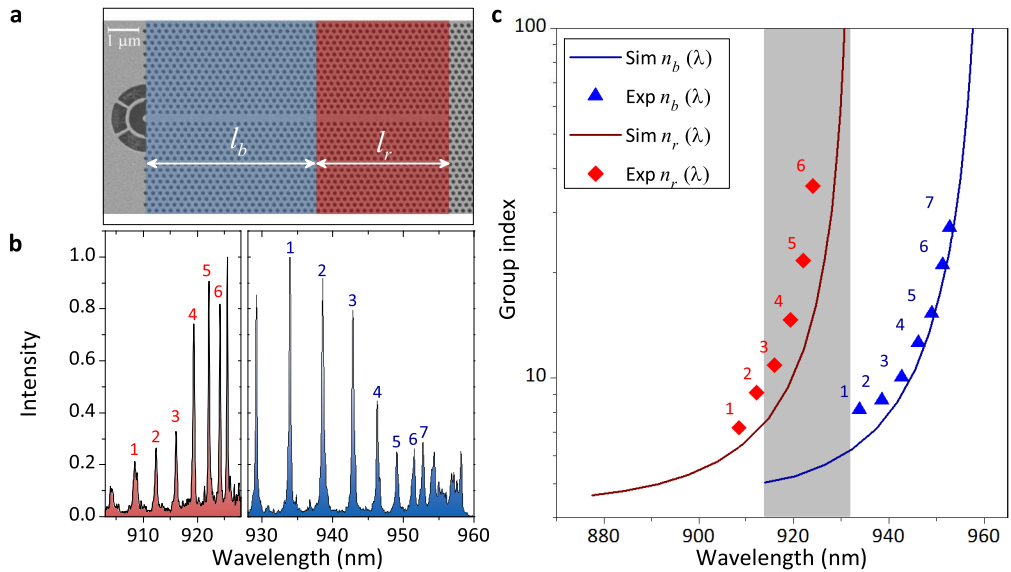}
\end{center}
\caption{(a) Scanning electron microscope image of the sample used to extract the group index. The waveguide sections supporting low-$n_\text{g}$ and high-$n_\text{g}$ modes in the wavelength region of interest are highlighted in blue and red, respectively. Their lengths $l_\text{b}$ and $l_\text{r}$ are indicated. (b) High-power spectra of the Fabry-P\'{e}rot resonances when exciting the two different sections of the sample. (c) Measured group index in the blue section (blue triangles) and in the red section (red diamonds), and corresponding theoretical values obtained from eigenvalue simulations (lines). The wavelength range of interest ($\simeq$ 910-930 nm) for the \textbeta-factor measurements is highlighted in grey.}\label{fig:Supp_Figure1}
\end{figure}

All waveguide samples used in the experiments are terminated by either inverse tapered or second-order Bragg grating mode adapters \cite{bib:Faraon.2008.S, bib:Tran.2009.S}. These mode adapters are best suited for coupling out low $n_\text{g}$ (fast propagating) modes. However, light-matter interaction is strongest for high $n_\text{g}$ (slowly propagating) modes. A typical sample combining two distinct waveguide sections is shown in Figure S\ref{fig:Supp_Figure1}(a). The two sections shown as the blue and red shaded regions are designed to support low-$n_\text{g}$ and high-$n_\text{g}$ modes, respectively, within their overlapping wavelength range ($\simeq$ 910-930 nm). The determination of the dispersion relation for the waveguides allowed for a detailed comparison of the measured \textbeta-factor with numerical simulations. For this purpose samples with increased reflectivity at the ends of the waveguide structure were fabricated in order to create sharp Fabry-P\'{e}rot resonances. The free spectral range, $\Delta\lambda_i = \lambda_i^2/2L$, is given by the center wavelength of the nearest peak, $\lambda_i$, and the optical path length between the two reflective surfaces, $L=l n_\text{g}$, where $l$ and $n_\text{g}$ denote the physical length of the structure and the group index.
A transition region between the red and the blue waveguide sections coupling the different $n_\text{g}$ modes is used in order to minimize reflections at these interfaces \cite{bib:Hugonin.2007.S}.
Figure S\ref{fig:Supp_Figure1}(a) shows an example of a sample that can be used to extract the dispersion relation of the waveguide mode. These waveguides are terminated by a photonic crystal on one side and by a grating mode adapter on the other side. While the waveguides have the same design parameters as the samples where the \textbeta-factors have been measured, the grating is designed for longer wavelengths and has a different duty cycle than the gratings used in \textbeta-factor measurements leading to an increased reflection back into the waveguide. The duty cycle is defined as the ratio between the optical length in the material and the optical length outside the material for a single period of the grating. The band edge of the red (blue) waveguide mode is spectrally positioned at around 930 nm (960 nm).
The measurements are performed by moving the excitation spot along the waveguide from the blue to the red section, and collecting the photoluminescence from the grating.

When the sample is excited in the blue section, the wavelengths above the band edge of the red waveguide mode can not propagate in the red section of the sample, therefore the interface behaves like a high-reflectivity mirror. Due to the high reflectivity of both the transition region and the grating structure, sharp Fabry-P\'{e}rot resonances appear in the spectral range between 930 and 960 nm. A high-power spectrum of these resonances is shown on the right side of Figure S\ref{fig:Supp_Figure1}(b). The wavelength dependent group index $n_\text{b}(\lambda)$ for the blue waveguide mode is extracted as \cite{bib:1.S}:
\begin{equation}
n_\text{b}(\lambda_i) = \frac{\lambda_i^2}{2l_\text{b} \Delta\lambda_i}
\end{equation}
where $\lambda_i$ is the central wavelength of the $i$th resonance, $l_\text{b}$ the length of the blue waveguide section, and $\Delta\lambda_i$ the free spectral range. The free spectral range around the resonance is approximated as the average of the free spectral range on either side of the resonance $\Delta \lambda_i = (\lambda_{i+1} - \lambda_{i-1})/2$.
The central wavelength $\lambda_i$ has been extracted by fitting each peak in the high-power spectrum with a Lorentzian lineshape. Due to the presence of Anderson localization in the proximity of the band edge, the analysis has been limited to the wavelength region in which the Fabry-P\'{e}rot resonances could be clearly distinguished from resonances arising due to random localization.

When the excitation spot is in the red waveguide section, only wavelengths below 930 nm can propagate to the grating, and the Fabry-P\'{e}rot fringes shown on the left side of Figure S\ref{fig:Supp_Figure1}(b) can be observed. In this second case, the optical path length is given by $L = n_\text{b}(\lambda) l_\text{b} + n_\text{r}(\lambda) l_\text{r}$, where $l_\text{r}$ is the length of the red waveguide section and $n_\text{r}(\lambda)$ its wavelength dependent group index. The group index $n_\text{r}(\lambda)$ can be calculated from:
\begin{equation}
\label{eq:FPs}
n_\text{r}(\lambda_i) = \frac{\lambda_i^2}{2l_\text{r} \Delta\lambda_i} - n_\text{b}(\lambda_i) \frac{l_\text{b}}{l_\text{r}}.
\end{equation}
The values of $n_\text{b}(\lambda)$ and $n_\text{r}(\lambda)$ extracted experimentally are shown in Figure S\ref{fig:Supp_Figure1}(c), together with the corresponding theory obtained from eigenvalue simulations of the structure. The calculated values of $n_\text{b}(\lambda)$ for $\lambda<930$ nm have been used to extract $n_\text{r}(\lambda)$ according to Eq. (\ref{eq:FPs}). For the specific waveguide shown here the group index reaches $n_\text{r}=33$. Furthermore we have measured $n_\text{g} > 50$ for several samples close to the band edge of the high-$n_\text{g}$ waveguide sections, which is where the high \textbeta-factors are observed in our experiments.

\subsection*{B. Optical measurements}

The QDs were pumped optically using a Ti:Sapphire laser, which provided 3-ps pulses at 800 nm with a repetition rate of 76 MHz. QDs with a slow decay rate were also investigated with a ps-pulse diode laser with a variable repetition rate operated at 20 and 40 MHz and an excitation wavelength of 785 nm.
The grating samples were investigated in a helium-flow cryostat operated at 10 K. Both excitation and collection were performed with the same microscope objective (NA $=$ 0.65), where the excitation beam was moved to the center of the waveguide while the collection was done from the grating.
The tapered samples were mounted in a helium-bath cryostat operating at 4.2 K. The excitation beam was focused by a top microscope objective with NA $=$ 0.5, and emission was collected from the taper with another microscope objective with NA $=$ 0.65. The sample and the excitation objective were positioned by two stacks of piezoelectric stages enabling motion in all three dimensions. Emission was sent to a spectrometer (600 or 1200 grooves/mm grating), where a CCD was used for spectral measurements and avalanche photodiodes were used to record either the decay dynamics of the emitters as well as their intensity autocorrelation function.

\subsection*{C. Comparison of measurement methods}
Different attempts of measuring the high \textbeta-factor in photonic-crystal waveguides have been reported in the literature. Lund-Hansen \emph{et al.} \cite{bib:3.S} reported $\beta = 89\%$ by the use of a top-collection confocal setup for both excitation and collection. With this method, however, only the photons leaking out of the waveguide are collected, which preferentially probes QDs with relatively low coupling efficiency since the high \textbeta-factor QDs emit out of plane very weakly and are thus not detectable. Furthermore, it is challenging to properly determine the decay rate of an uncoupled QD by top-collection measurements, since the spatial resolution of the collection optics is not sufficiently high to ensure that all detected lines originate from QDs spatially positioned in the waveguide.
Finally, in this work the QD lines used to extract $\Gamma_{\text{uc}}$ are spectrally positioned in the band gap of the photonic crystal, where the coupling to radiation modes is weaker than in the waveguide mode \cite{bib:Lecamp.2007.S}. This can potentially lead to overestimates of the \textbeta-factor.

In order to overcome the potential issue of spatial mismatch, one approach has been to temperature tune a Purcell-enhanced QD across the waveguide band edge and into the band-gap region \cite{bib:4.S}. However, this approach is limited by the rather elevated temperatures required to obtain a sufficiently large tuning range, which cause an increase in the nonradiative decay $\Gamma_{\text{nr}}$ and potentially residual coupling to the waveguide mode due to phonon-mediated processes \cite{bib:2.S}. Both these effects can significantly contribute to $\Gamma_{\text{uc}}$, and thus limited the extracted values of $\beta$ to 85.4\%.
In recent work, experimental efforts have shifted towards the direct collection of photons coupled to the waveguide mode \cite{bib:5.S, bib:6.S, bib:7.S}. However, $\Gamma_{\text{uc}}$ was still extracted in a top-collection scheme \cite{bib:5.S, bib:6.S}, either from QDs sitting in a defect-free PC or with a method similar to that used in Ref. \cite{bib:3.S}. These approaches have the drawback of neglecting the changes of the coupling to the radiation modes due to the presence of the line defect. In addition Ref. \cite{bib:7.S} did not take into account any variation of $\Gamma_{\text{nr}}$ between different QD samples used in different experiments.

\subsection*{D. $g^{(2)}(\tau)$ measurements}

Measurements of the second order intensity correlation function were performed to ensure the single-photon nature of the emission from the QDs. Due to the strong Purcell enhancement in proximity of the waveguide band edge and the the use of a density of QDs of ~100 \textmu m$^{-2}$, a pronounced background emission is present in high power spectra. For this reason, precisely pinpointing the saturation power for single QD lines by mean of intensity measurements was not straightforward, and only an estimate of the magnitude of the saturation power could be obtained from decay curve analysis. The approximate knowledge of the saturation power was used to ensure that time resolved and intensity correlation measurements under above band excitation were taken at powers well below saturation.
\begin{figure}
\begin{center}
\includegraphics[scale = 1]{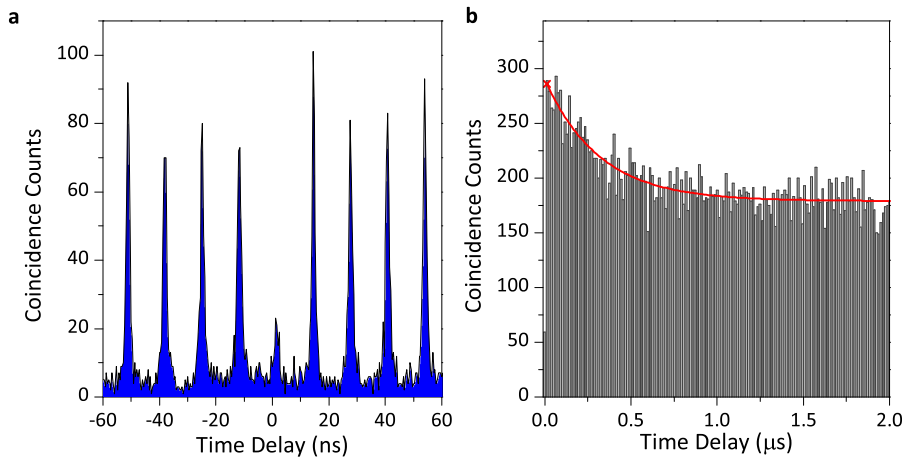}
\end{center}
\caption{(a) Intensity correlation function from a Purcell enhanced quantum dot, recorded under above-band excitation (800 nm) and at power well below saturation. (b) Intensity correlation histogram over a 2 \textmu s time window, recorded under longer-wavelength excitation (corresponding to the function in Figure 2(i) in the main text). The red line indicates the single exponential fit to the data.}\label{fig:Supp_Figure4}
\end{figure}
An example of a $g^{\text{(2)}}(\tau)$ measurement recorded with above-band excitation (800 nm) at low power is shown in Figure S\ref{fig:Supp_Figure4}(a). The \textbeta-factor of the QD in this case was $\beta =$ 94\%.

In order to obtain cleaner spectra and measure the saturation power precisely, excitation at longer wavelengths was performed. The data presented in Figure 2(i) of the main text have been recorded by exciting the QD at 880.02 nm. The excitation power was fixed to 0.63 times the saturation power of the quantum dot transition.
A power series was used to extract the saturation power and to confirm that the line under investigation corresponds to a single exciton line.
The decay rate of the QD measured under higher-wavelength excitation was $\Gamma_{\text{c}}$ = 6.28$\pm$0.15 ns$^{-1}$, corresponding to $\beta = $98.43$\pm$0.04\%.

Figure S\ref{fig:Supp_Figure4}(b) shows the autocorrelation function presented in Figure 2(i) of the main text for time delays of up to 2 \textmu s. One can notice the bunching at time delays close to zero, which is the typical signature of blinking \cite{bib:Santori.2004.S}. Since the relevant time scale for operation of the single-photon source is the QD lifetime, we compare the residual central peak to its first neighbouring peak \cite{bib:Vuckovic.2003.S}. We obtain $g^{\text{(2)}}(0)$ = 0.2, which quantifies the amount of multi-photon emission of the source. The area of the first neighbouring peak has been extracted by fitting the data shown in Figure S\ref{fig:Supp_Figure4}(b) with a single exponential decay, and evaluating the fit at $\tau$ = 13 ns.
Following the method presented in Ref. \cite{bib:Santori.2004.S}, we extract an excitation efficiency of ~40\%, which specifies the probability that the QD is emitting on the studied transition rather than blinking to another transition.

\subsection*{E. Simulation methods and results}

Numerical simulations using a finite-element method (FEM) in the frequency domain have been performed to model the coupling to radiation modes for a dipole emitter coupled to a PCW. The total power emitted by the dipole in the waveguide relative to the total emitted power in a homogeneous medium can be expressed as $P/P_{\text{hom}}=(\Gamma_{\text{wg}} +\Gamma_{\text{rad}})/\Gamma_{\text{hom}}$, where $\Gamma_{\text{hom}} = $ 0.91$\pm$0.08 ns$^{-1}$ is the radiative decay rate of the QD in a homogeneous medium that is recorded experimentally.  By computing the values of $\Gamma_{\text{wg}}$ and $\Gamma_{\text{rad}}$ (see below) for specific positions and dipole orientations, the radiative \textbeta-factor can be obtained from $\beta = \Gamma_{\text{wg}}/(\Gamma_{\text{wg}} + \Gamma_{\text{rad}})$, including the experimentally extracted value of $\Gamma_{\text{nr}}$ leads to Figure 1(e) in the main text.

The problem of defining appropriate boundary conditions for the waveguide mode in the numerical simulations has been overcome by applying a set of active boundary conditions at the two ends of the waveguide \cite{bib:Lecamp.2007.S, bib:Chen.2010.S}. The active boundary conditions absorb the propagating field with a mode profile given by the Bloch mode and a field amplitude extracted from eigenvalue simulations. Perfectly matched layers (PMLs) are used at all the boundaries of the simulation domain except for the two planes at the ends of the PCW, where active boundary conditions are applied. The total power radiated by the dipole, $P$, is extracted by integrating the power flowing out of a small cube surrounding the emitter. The power coupling to radiation modes, $P_{\text{rad}}$, is extracted with the same method, but using a box that surrounds the simulation domain. The yz-faces of this second box were left open to avoid including any contribution from the waveguide mode in the integration. Convergence tests have been performed to ensure the accuracy of the results with respect to mesh size, length of the waveguide, height of the active boundary conditions, and size of the box used for the power integration. The power emitted into the waveguide mode can be then extracted from $P_{\text{wg}}=P-P_{\text{rad}}$.

Using the above outlined method $\Gamma_\text{rad}$ is calculated for both x and y dipoles at different spatial positions within the waveguide and the results are presented in Figure 4(a) in the main text. These calculations are performed for $n_\text{g} = 58$ and for $n_\text{g} = 7$ in structures corresponding to the high-$n_\text{g}$ and to the low-$n_\text{g}$ waveguide sections, respectively. In Figure 4(b) in the main text, $\Gamma_{\text{wg}}+\Gamma_\text{rad}$ is shown for the same dipoles. The spectral positions within the respective waveguides were chosen to check the validity of the method used to extract the high \textbeta-factors from the measurements of QDs close to the band edge of the high-$n_\text{g}$ waveguide mode. An equivalent investigation was performed for a dipole far blue detuned from the waveguide mode band edge, $n_\text{g} = 5$, in structures corresponding to the high-$n_\text{g}$ waveguide section. These simulations where used to show the validity of our method for the QD detuned 20nm from the waveguide mode band edge and still exhibiting a \textbeta-factor above 90\%. Finally the two simulations performed at $n_\text{g} = 58$ and $n_\text{g} = 5$ were combined to show the broadband nature of these structures, which is presented in Figure 1(e) of the main text.

\subsection*{F. Decay curve analysis}

\begin{figure}
\begin{center}
\includegraphics[scale = 1]{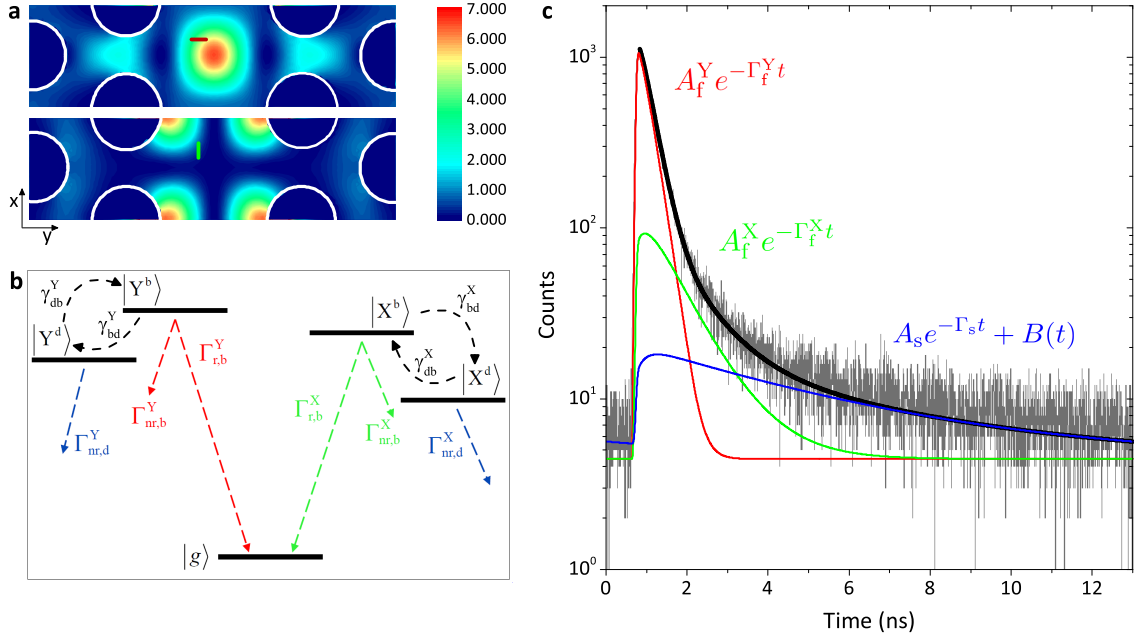}
\end{center}
\caption{(a) Spatial map of the calculated Purcell factor within the waveguide unit cell at $n_\text{g}=36$ for (top) a y-polarized and (bottom) an x-polarized dipole. The red (green) bar represents a QD aligned along the y- (x-)direction and situated at a point where both the bright exciton states of the QD interact with the Bloch mode. (b) Five level scheme of the first excited state of a QD. The main decay channels for each state are indicated. (c) Decay curve fitted with a triple exponential model. The black curve is the total fit to the data, whereas the red, green, and blue curves indicate the two fast decay components and the slow component, respectively.}\label{fig:Supp_Figure3}
\end{figure}

The lowest exciton state of an InAs QD is described by a five-level model, formed by the ground state, two bright states, and two dark states \cite{bib:10.S}. Due to an anisotropy in the strain, shape, and composition of the QD, the exchange interaction separates the bright states in energy by the fine structure splitting. These states become linearly polarized along the lattice directions $[110]$ and $[1\overline{1}0]$ corresponding to the x- and y-direction of the photonic-crystal lattice shown in Figure S\ref{fig:Supp_Figure3}(a) \cite{bib:Bennett.S}. For this reason, the exciton states are labeled respectively $|\text{X}\rangle$ and $|\text{Y}\rangle$.

The QD energy diagram is shown in Figure S\ref{fig:Supp_Figure3}(b) together with an indication of the main decay channels for each energy level.
The two bright exciton states $|\text{X}^\text{b}\rangle$ and $|\text{Y}^\text{b}\rangle$ can decay either radiatively to the ground state with decay rate $\Gamma_{\text{r,b}}$ or nonradiatively ($\Gamma_{\text{nr,b}}$), or decay into a dark state through a spin-flip process ($\gamma_{\text{bd}}$). The decay rate of the bright state can thus be expressed as the fast decay rate $\Gamma_{\text{f}}^{\text{X,Y}} = \Gamma_{\text{r,b}}^{\text{X,Y}} + \Gamma_{\text{nr,b}} + \gamma_{\text{bd}}$. The dark states $|\text{X}^\text{d}\rangle$ and $|\text{Y}^\text{d}\rangle$ can not recombine through the emission of a photon, so they decay only nonradiatively to the ground state or into the bright state via spin flip with the slow decay rate $\Gamma_{\text{s}}^{\text{X,Y}} =\Gamma_{\text{nr,d}} + \gamma_{\text{db}}$. At the relevant temperatures for the present experiment $\gamma_{\text{db}} = \gamma_{\text{bd}}.$
The bright-bright spin-flip rates are neglected in the model, since they are much smaller than the other decay rates \cite{bib:Wang.2011.S}. For this reason, the $|\text{X}\rangle$ and $|\text{Y}\rangle$ excitons can be treated independently.

When the QD is embedded in a homogeneous medium, the model describing the dynamics of the quantum dot population is a bi-exponential decay, $\rho_{\text{e}}(t) = A_{\text{f}} e^{-\Gamma_{\text{f}} t} + A_{\text{s}} e^{-\Gamma_{\text{s}} t}$, where $\Gamma_{\text{f}} = \Gamma_{\text{f}}^{\text{X}} \simeq \Gamma_{\text{f}}^{\text{Y}}$, $\Gamma_{\text{s}} = \Gamma_{\text{s}}^{\text{X}} \simeq \Gamma_{\text{s}}^{\text{Y}}$, and $A_{\text{f}}$ and $A_{\text{s}}$ are the amplitudes of each decay component. This model is derived by solving the rate equations for the population of each energy level \cite{bib:Wang.2011.S}.
For QDs in a photonic-crystal waveguide, the coupling to the waveguide mode changes the dynamics of the emitter. Figure S\ref{fig:Supp_Figure3}(a) shows the Purcell factor distribution at $n_\text{g} = 36$ for a dipole oriented along the y-direction (top) and the x-direction (bottom). For certain positions in the waveguide unit cell, such as the one indicated in the figure, both orthogonal dipoles $|\text{X}\rangle$ and $|\text{Y}\rangle$ are coupled to the waveguide mode. However, the dipole oriented along the y-direction experiences a different Purcell enhancement than the dipole oriented along the x-direction.
In this case, the decay curves are triple exponentials:
\begin{equation}
\rho_\text{e}(t) = A_{\text{f}}^{\text{Y}} e^{-\Gamma_{\text{f}}^{\text{Y}} t} + A_{\text{f}}^{\text{X}} e^{-\Gamma_{\text{f}}^{\text{X}} t} + A_{\text{s}} e^{-\Gamma_{\text{s}} t} + B(t)
\end{equation}
where $B(t)$ accounts for a time dependent background level due to the contribution of other emitters coupled to the waveguide mode. This background level is especially pronounced in high-density samples, where it may dominate the contribution from $A_{\text{s}} e^{-\Gamma_{\text{s}} t}$.
The example of a decay curve fitted with the triple exponential model is shown in Figure S\ref{fig:Supp_Figure3}(c), together with the individual decay components of the fit. The total radiative decay rate is given by the sum of the decay rate into the waveguide mode and the one into radiation modes ($\Gamma_{\text{r}} = \Gamma_{\text{wg,b}}+\Gamma_{\text{rad,b}}$). Since it has been shown experimentally that $\Gamma_{\text{nr,b}} = \Gamma_{\text{nr,d}}$ \cite{bib:Johansen.2010.S}, the radiative decay rate can be determined as $\Gamma_{\text{f}} - \Gamma_{\text{s}}.$

\subsection*{G. Single-photon source efficiency}

\begin{figure}
\begin{center}
\includegraphics[scale = 1]{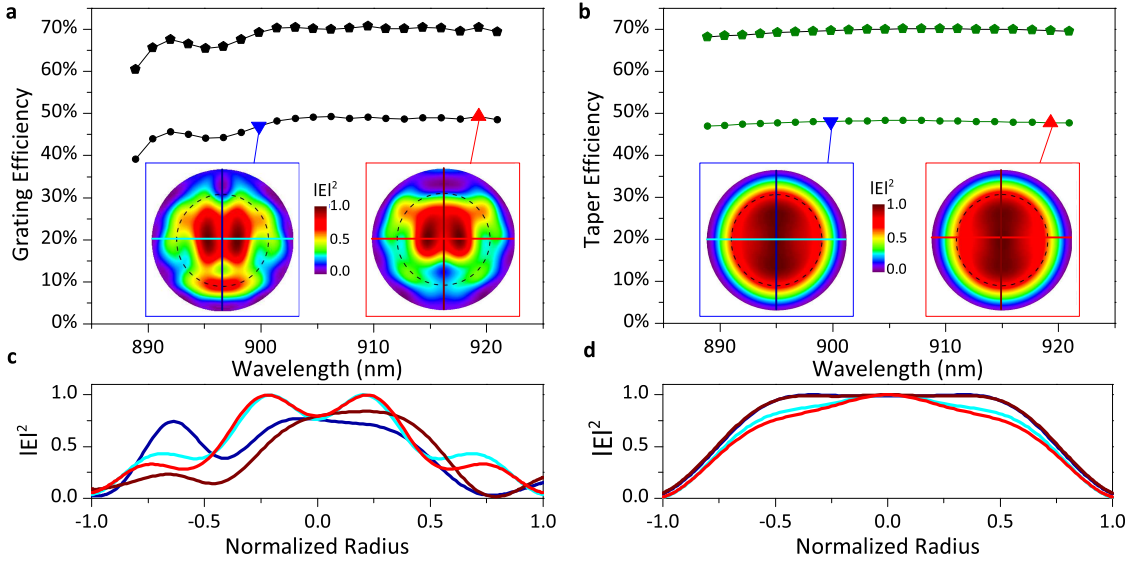}
\end{center}
\caption{(a)-(b) Collection efficiencies and far-fields for samples with (a) grating outcouplers, and (b) tapers. The plots show the collection efficiency at a lens with $\textrm{NA} = 0.65$ (circles) and $\textrm{NA} = 0.82$ (diamonds) as a function of wavelength along the waveguide band. The insets show the far-field radiation patterns at two different wavelengths with the circle indicating the angles that fall within the numerical aperture of the lens used in the experiment ($\textrm{NA} = 0.65$). The centre of the far-field patterns correspond to vertical emission, while the edge correspond to emission towards the horizon. (c)-(d) Cross-sections of the far-field patterns. The colors correspond to the lines indicated in the insets of (a) and (b).}\label{fig:Supp_Figure5}
\end{figure}

We analysed the collection efficiency of the inverse taper and second-order Bragg grating mode adapters using finite-element frequency domain calculations. In these calculations, a dipole source is placed in the waveguide and its emission couples into the waveguide mode. As previously mentioned, the waveguides contain a high $n_g$ region followed by a low $n_g$ region, which is then connected to the out-coupler.
We compute the fraction of light collected by a lens of given numerical aperture at one of the outcouplers. The results for the second-order Bragg grating is shown in Figure S\ref{fig:Supp_Figure5}(a). The gratings are optimized for a wavelength of $920$ nm which approximately corresponds to the band-edge of the waveguide. We computed that $\sim$45\% of the light is collected by a lens with a numerical aperture $\textrm{NA}=0.65$, corresponding to the one used in the experiment, and that $\sim$70\% of the emission can be collected by using a higher numerical aperture ($\textrm{NA}=0.82$). 

Our samples have outcouplers on both sides of the waveguide, and each grating couples light out of the waveguide mode to two different directions (up or down). Therefore the ratio of the light collected at the lens to the total power in the waveguide mode corresponds to 25\% of the efficiency shown in Figure S\ref{fig:Supp_Figure5}(a). The insets in Figure S\ref{fig:Supp_Figure5}(a) show the far-field distribution intensity of light emitted from the gratings at two different wavelengths. As the far-fields are non-Gaussian, they are not generally expected to couple well to single mode fibres although this has not been quantified further. In order to better clarify this point, two orthogonal cross sections of the far-field patterns are shown in Figure S\ref{fig:Supp_Figure5}(c).  Figure S\ref{fig:Supp_Figure5}(b) shows the efficiency calculations for the inverse taper mode adapters. These structures have an efficiency of $\sim$45\% across the waveguide band for a lens with $\textrm{NA}=0.65$ and up to $\sim$70\% for a lens with $\textrm{NA}=0.82$.
Since the mode can be coupled out on both sides of the waveguide, only half of the light is in the waveguide is coupled by a single taper. The far-field intensity shown in the insets of Figure S\ref{fig:Supp_Figure5}(b) indicates that the tapers have a more symmetric emission pattern than the gratings, although the cross sections of the far-field pattern shown in Figure S\ref{fig:Supp_Figure5}(d) reveal that the mode is non-Gaussian, which will influence the coupling to a single-mode fiber.

The outcoupling efficiency of the gratings and tapers can be used to estimate the highest achievable count rate on the APD.
In the limit where the quantum dot emits a single photon each time it is triggered by an optical pulse, the figure of merit for the maximum number of photons available inside the waveguide is the beta-factor.
Since the repetition rate of the laser in the experiment was 76 MHz, this leads to 74.8 million photons per second created in the photonic-crystal waveguide for a beta-factor of 98.4\%. Since the propagation losses in our structures are negligible, this number is not significantly reduced after propagation. The presence of blinking observed in the autocorrelation measurement translates to an effective reduction of the excitation efficiency of the QD transition, which reduces the count rate in the waveguide to $\sim$30 million per second; however electrically gated structures combined with resonant excitation have proven capable of eliminating this issue \cite{bib:Simon.2011.S}.
The expected count rate at the photon detector can be estimated by considering the collection efficiency in the first lens of the outcouplers, and estimating the transmission efficiencies of each optical component in the setup. We have a beam sampler with $\sim$90\% transmission in the collection arm, a total $\sim$50\% loss in the fiber-to-fiber mating sleeves, 20\% throughput of the spectrometer, and 35\% efficiency of the single photon detector, which lead to an overall setup collection efficiency of $\sim$3\%. Putting these numbers together leads to a highest achievable count rate of $\sim$100.000 counts per second from the grating structures. In our experiment, we typically measure $\sim$10.000 counts per second. The discrepancy between expected and measured count rates is most likely due to the mismatch between the non-Gaussian shape of the outcoupled mode and the single-mode fiber mode. This could straightforwardly be improved by further design of the outcoupling structure.

\end{document}